\documentclass[%
reprint,
 unsortedaddress,
 amsmath,amssymb,
 aps,prb
]{revtex4-1}

\usepackage{amsmath,amssymb}
\usepackage{euscript} % for \EuScript{}
\usepackage{mathptmx} % Use Times as default text font
\usepackage{ulem} % Package for underlining and strikeout
\usepackage{graphicx}% Include figure files
\usepackage{dcolumn}% Align table columns on decimal point
\usepackage{bm}% bold greek math
\usepackage{braket}% bra ket notation
\usepackage{pifont}%circled number
\usepackage[usenames, dvipsnames]{color}
\usepackage{float}
\usepackage{hyperref}% add hypertext capabilities
\usepackage[mathlines]{lineno}% Enable numbering of text and display math
\mathchardef\mhyphen="2D %define hyphen in math

\begin{document}

\preprint{APS/123-QED}

\title{Weyl points and topological surface states in a three-dimensional elastic lattice}
\author{Sai Sanjit Ganti}
\author{Ting-Wei Liu}
\author{Fabio Semperlotti}
 \email{fsemperl@purdue.edu}
\affiliation{Ray W. Herrick Laboratories, School of Mechanical Engineering, Purdue University, West Lafayette, Indiana 47907, USA.}

%\date{ }

\begin{abstract}
%Weyl semimetals are quantum electronic materials whose dispersion is characterized by three-dimensional (3D) nodal degenerate points called the Weyl points. The nontrivial topological nature of these points allows the material to support unidirectional 3D surface modes along the material boundaries that are immune to back scattering due to disorder or defects.
Following the realization of Weyl semimetals in quantum electronic materials, classical wave analogues of Weyl materials have also been theorized and experimentally demonstrated in photonics and acoustics. Weyl points in elastic systems, however, have been a much more recent discovery. In this study, we report on the design of an elastic fully-continuum three-dimensional material that, while offering structural and load-bearing functionalities, is also capable of Weyl degeneracies and surface topologically-protected modes in a way completely analogous to the quantum mechanical counterpart. The topological characteristics of the lattice are obtained by \textit{ab initio} numerical calculations without employing any further simplifications. The results clearly characterize the topological structure of the Weyl points and are in full agreement with the expectations of surface topological modes. Finally, full field numerical simulations are used to confirm the existence of surface states and to illustrate their extreme robustness towards lattice disorder and defects. 
\end{abstract}

\maketitle

\section{Introduction}
\label{sec1}
%\FS{TW: cite your works on QVHE and QSHE where appropriate.}
The search for novel topological states of matter has recently seen many exciting breakthroughs either in quantum or classical wave physics \cite{ren,hasan,lu2014,zhang2018,ma2019topological}. Phenomena such as the quantum Hall effect (QHE), the quantum spin Hall effect (QSHE), and the quantum valley Hall effect (QVHE) have been studied extensively and their nontrivial topological characteristics have been clearly connected to their ability to support back-scattering-immune topological edge states \cite{ren,hasan}. Although the concept of topological material was discovered and developed in the field of condensed matter physics, the many conceptual and mathematical similarities between different fields of wave physics have led to the development of analogue concepts of topological materials in electromagnetic, acoustic, and elastic systems \cite{ValleySonicBulk,ValleySonicEdge,pal2017edge,vila2017observation,yan2018chip,snowflake,lu2014,liu2018tunable,zhu2018design,liu2019experimental,liu2019nonconventional,zhang2018,ma2019topological,acousticTI2017,zonefold2,kekuleprl2017,kekulenjp2018,zonefold2018,MechanicalTI,AcousticTIPlate,AcousticTIAir,miniaci2018experimental,chen2019mechanical}. Recent studies have identified the existence of the so-called Weyl semimetals as three dimensional nontrivial topological materials capable of unidirectional back-scattering-protected surface states \cite{yan2017topological}.

Weyl semimetals are quantum materials whose behavior is characterized in terms of the Weyl Hamiltonian $H(\mathbf{k})=\nu_xk_x\sigma_x+\nu_yk_y\sigma_y+\nu_zk_z\sigma_z$, where $\nu_i$, $k_i$, and $\sigma_i$ represent the components of the group velocity, momentum, and Pauli matrix, respectively. These materials can be thought of as a three-dimensional extension of two-dimensional nontrivial topological materials characterized by Dirac degeneracies such as, for example, systems exhibiting QHE, QSHE, and QVHE. Similar to 2D Dirac materials, Weyl semimetals possess a degenerate nodal point formed by linear intersection of two bands in the three coordinate directions of the three dimensional reciprocal space. This degenerate point, called the Weyl point, carries a nonzero topological charge which confers the material nontrivial topological properties. The charge (i.e. the integral of the Berry curvature flux in a 2D manifold enclosing the Weyl point) can have either a positive or negative sign, hence determining if the degeneracy acts as a quantized source or a sink of berry curvature flux% and therefore are treated as an equivalent of magnetic monopoles in reciprocal space
. The total topological charge of a Brillouin zone (BZ) in reciprocal space should always be zero \cite{nielsen1981absence} and therefore Weyl points always occur in pairs of opposite charge. Nontrivial topological surface states connecting the Weyl points of opposite charge exist on the boundary of the system. Apart from topological surface states, Weyl points have also been linked to other unusual effects such as chiral anomaly \cite{nielsen1983adler} and quantum anomalous Hall effect \cite{yang2011quantum}.

Weyl points can be conceptually interpreted as the  extension of Dirac points to three-dimensional lattices. However, some fundamental differences exist that have important implications on the robustness of these degeneracies. The existence of Dirac points can be guaranteed based on symmetry arguments (e.g. graphene-like lattices), provided that both parity-inversion ($P$) and time-reversal ($T$) symmetries are preserved. Breaking either $P$ or $T$-symmetry would lift the degeneracy leaving behind a bandgap with topological significance.
Unfortunately, the existence of Weyl points is much more ambiguous and it cannot be guaranteed \textit{a priori}. However, some useful indications can come from the analysis of symmetries. In the case of Weyl points, $P$ and $T$-symmetries impose contradicting requirements. In fact, considering a lattice having a Weyl point at the $\mathbf{k}$ high-symmetry point in reciprocal space, $T$-symmetry would require the topological charges at $\mathbf{k}$ and $-\mathbf{k}$ to be of the same sign while $P$-symmetry would require them to be opposite. Due to these contradicting requirements, it follows that Weyl points can only be obtained in systems with broken \textit{PT}-symmetry. Thus, unlike for Dirac points in which symmetry breaking is a necessary condition to lift the degeneracy, for Weyl points symmetry breaking is a requirement for the existence of the isolated degeneracy. These more stringent requirements on symmetry conditions are at the foundation of the higher robustness of Weyl points against external perturbations. In fact, previous studies showed that they can only be annihilated by combining pair of points with opposite charges. %Therefore, surface states obtained in Weyl materials are robust to perturbations as long as the translation symmetry is preserved. 
Also, following the requirements dictated by both $P$ and $T$-symmetries, the minimum number of Weyl points in a BZ is 4 (2 pairs) when $P$-symmetry is broken and 2 (1 pair) when $T$-symmetry is broken \cite{lu2014}.

Note also that, although in the current study only Weyl points with unit charge are considered, other studies \cite{xu2015discovery,lv2015experimental,xu2015,fang2012multi,chen2016photonic,wang2018multiple,chang2017multiple,chen2018acoustic,liu2018acoustic} have reported systems possessing Weyl points having higher charge  typically caused by the overlapping of multiple Weyl points of unit charge. Such points can be readily identified from the band structure because the modes do not intersect linearly at these points in the $k_x\mhyphen k_y$ plane. However, the intersection along $k_z$ would still be linear. %A higher topological charge indicates that the Weyl point is more robust to perturbations. 
Recent studies have also reported a new kind of Weyl point labeled type-II both in photonic and phononic systems \cite{soluyanov2015type,xiao2016hyperbolic,yang2017direct,noh2017experimental,yang2016acoustic,xie2019experimental}. The degeneracies discussed in the present study are of type-I and represent the more direct analogue to the degenaracies originally studied by Weyl in quantum mechanical systems \cite{weyl1929elektron}.

While first theorized as the massless solution of the Dirac equation \cite{weyl1929elektron}, Weyl fermions could be realized and experimentally demonstrated in Weyl semimetals only recently \cite{wan2011topological,xu2015discovery,lv2015experimental,xu2015,fang2012multi,singh2012topological,bulmash2014prediction,huang2015weyl,weng2015weyl}. The possibility of identifying classical mechanical systems serving as analogue to the Weyl semimetals has opened an intriguing and fertile topic of research spanning many areas of wave physics. As a result, non-quantum-mechanical analogues of the Weyl points were formulated both in photonic \cite{lu2015experimental,lu2013weyl,chen2016photonic,chang2017multiple,gao2016photonic,xiao2016hyperbolic,yang2017direct,noh2017experimental,yang2018ideal,bravo2015weyl,takahashi2018circularly,fruchart2018soft} and acoustic \cite{fruchart2018soft,li2018weyl,xiao2015synthetic,yang2016acoustic,ge2018experimental,chen2018acoustic,liu2018acoustic} metamaterial systems. Weyl points were realized and experimentally demonstrated in photonics using double-gyroid structure characterized by broken $P$-symmetry \cite{lu2013weyl,lu2015experimental}. Although successful in realizing Weyl points, the double-gyroid structure involves a complex fabrication. Simpler designs based, as an example, on woodpile photonic crystals \cite{takahashi2018circularly,chang2017multiple} or on rotated stacked rods in acoustic systems \cite{chen2018acoustic,liu2018acoustic} were also explored. A planar fabrication methodology, which outlines a step by step approach to obtain Weyl points, was also presented \cite{chen2016photonic,bravo2015weyl}.   

Despite the successful implementation of Weyl points in photonic and acoustic systems, the realization of Weyl points in elastic systems has proven to be more challenging than other classical analogues. This increased complexity stemmed from wave coupling and mode conversion between different polarizations as well as from the difficulty in fabricating the necessary 3D structures. To-date, successful realizations were reported only in a few studies \cite{wang2018multiple,shi2019elastic}. One study made use of beams and thin plates with a particular geometry \cite{wang2018multiple} purposely selected to be manufactured by additive manufacturing. Another study utilized a truss-like construction made of beam elements \cite{shi2019elastic}. Although elegant and simple to fabricate, these designs are discontinuous in nature therefore not well suited for applications where the load-bearing capabilities play a critical role (e.g. applications to structural materials).
Also, in both the above mentioned studies, the topological characteristics were obtained based on a tight-binding (TB) formulation of the Hamiltonian. This approach works well only when the fundamental lattice is composed of weakly coupled resonant elements. In addition, the hopping parameters used in the TB model usually cannot be clearly connected to the actual geometric and design parameters. These limitations raise the more important question of how accurately TB models can describe continuous elastic lattices.
In this regard, we propose an elastic analogue of a Weyl semimetal material based on a fully continuous design resulting in a solid load-bearing structure. 
Also, due to the strongly coupled nature of the elemnts in this design, we present a detailed study of the topological characteristics of such medium based on \textit{ab initio} calculations, hence bypassing the limitations of TB approaches.

This paper is organized as follows: Sec.~\ref{sec2} will introduce the proposed design and the corresponding dispersion properties in connection with Weyl points. Sec.~\ref{sec3} will present \textit{ab initio} calculations of the topological invariant. Sec.~\ref{sec4} will focus on the dispersion behavior of a supercell and the corresponding dispersion structure, in order to determine the occurrence of surface states. Finally, Sec.~\ref{sec5} will show full field simulations of the elastic Weyl material in order to illustrate the one-way, backscattering-immune nature of the surface states.
\section{Synthesis of the unit cell and dispersion properties}
\label{sec2}
From a general perspective, in order to develop the fundamental unit cell, we select a
%\ed{[I have discussed w/ prof. and he agreed keeping "triangular lattice"]} \FS{shouldn't we just say hexagonal here given also that all the vertices have same properties?}\SG{[Hexagonal refers more to the graphene like structure, where the cylinders would be present only at the vertices of the hexagon. In our case however, we have a cylinder even at the center of the hexagon and therefore this would be called a triangular lattice.]} \FS{Are you sure? I can see how a lattice with cylinders only at the vertices can be decomposed in two triangular sublattices but if you have a center cylinder how do you do that?}
2D triangular lattice (assumed in the $xy$-plane) and build the 3D geometry by periodically repeating this unit along the $z$-direction. The final result is a layered 3D structure having intact $P$ and $T$-symmetries. Note that, as previously mentioned, starting from a 2D triangular lattice guarantees (due to symmetry arguments) the existence of Dirac degeneracy at the high symmetry points in the $k_x\mhyphen k_y$ momentum space. Periodically repeating this unit in the $z$-direction extends the Dirac degeneracy at virtually any $k_z$ value, thereby generating a line node degeneracy \cite{lu2014,bravo2015weyl} along $k_z$. In order to lift this three-dimensional degeneracy, 
%\ed{[But why  at the bottom of page 2 ``Weyl Hamiltonian requires all the symmetries to be broken...''? I'd  suggest to remove ``all the'']}\FS{I agree. I think SG meant mirror and inversion.....I specified that. See if you guys agree.}\SG{[I agree Ting-Wei. I suggested a replacement for the said statement in section 1.  ]}
$P$-symmetry can be broken by the proper introduction of additional structural element that do not respect inversion symmetry. For this purpose, we select slanted beam elements connecting the vertices of the hexagonal structure on two adjacent layers as shown in  Fig.~\ref{fig:3}. The result is the introduction of a chiral coupling between the layers 
that breaks $P$-symmetry and all existing mirror symmetry. 
The most direct consequence is that the line degeneracy along $k_z$ is reduced to a point, hence giving rise to the Weyl point.  

In the following section, we first describe the geometry and the dynamic properties of the parity-preserving fundamental unit. Then, we introduce the chiral coupling and investigate its effect compared with the \textit{P}-symmetric unit.

\subsection{3D lattice with intact \textit{P}-symmetry: line node degeneracy}
\begin{figure*}[ht]
	\centering
	\includegraphics[scale=0.666]{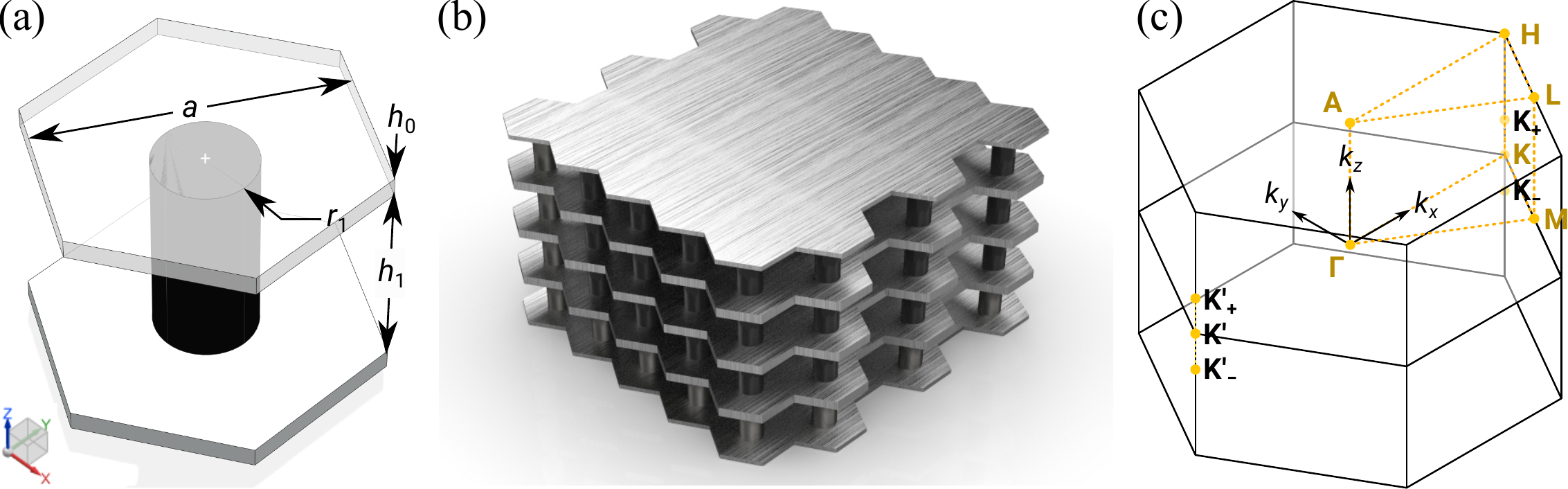}
	\caption{\label{fig:1}
	(a) Schematics of the unit cell with intact $P$- and $T$-symmetry. Some characteristic dimensions are shown. The top and bottom plates are made of aluminum (transparent white) while the cylinder is made of iron (black). Periodic boundary conditions are used on the boundaries to create a 3D lattice.
	(b) Rendered view of a bulk piece of the 3D lattice.
	(c) Brillouin zone of the resulting 3D lattice in reciprocal space. Orange dashed lines mark the irreducible part of the Brillouin zone (IBZ).}
\end{figure*}

Fig.~\ref{fig:1} (a) and (b) show the fundamental unit cell used to create the lattice. The cell is made of a vertical cylinder included between two homogeneous, hexagonal-shaped, thin plates. The cylinder has its longitudinal axis aligned with the $z$-direction. The cylinder is made of iron while the plates are made of aluminum. The selection of these materials was motivated by the intent of obtaining marked degeneracies at the high symmetry points in the momentum space. Clearly, several other combinations of structural materials leading to similar conditions could potentially be identified. The dimensions of the unit cell are: $a=30$ mm, $h_0=2$ mm, $r_1=5$ mm, $h_1=20$ mm, where $a$ is the lattice constant, $h_0$ is the thickness of each homogeneous flat plate, $r_1$ and $h_1$ are the radius and height of the cylinder, respectively. A 2D lattice in the $x$-$y$ plane can be assembled by periodically repeating the unit cell. The final result is a lattice of cylinders in a hexagonal configuration sandwiched between two thin plates. This 2D lattice can be periodically repeated in the $z$-direction to form the final 3D lattice.

The analysis of the band structure of the 3D material can be conveniently performed by applying periodic boundary conditions to the fundamental unit cell. More specifically, periodic boundary conditions can be applied along the sideward faces of the plate in order to create the 2D lattice in the $xy$-plane, while periodic boundary conditions applied on the top and bottom faces of the top and bottom plates will yield the complete 3D lattice.
The resulting system is characterized by intact $P$ and $T$-symmetries and the corresponding 3D Brillouin zone is shown in Fig.~\ref{fig:1} (c). For a fixed value of the momentum component $k_z$, the 3D BZ simplifies into a 2D BZ typical of a triangular lattice.
% Figure 1

\begin{figure*}[ht]
	\centering
	\includegraphics[scale=0.33]{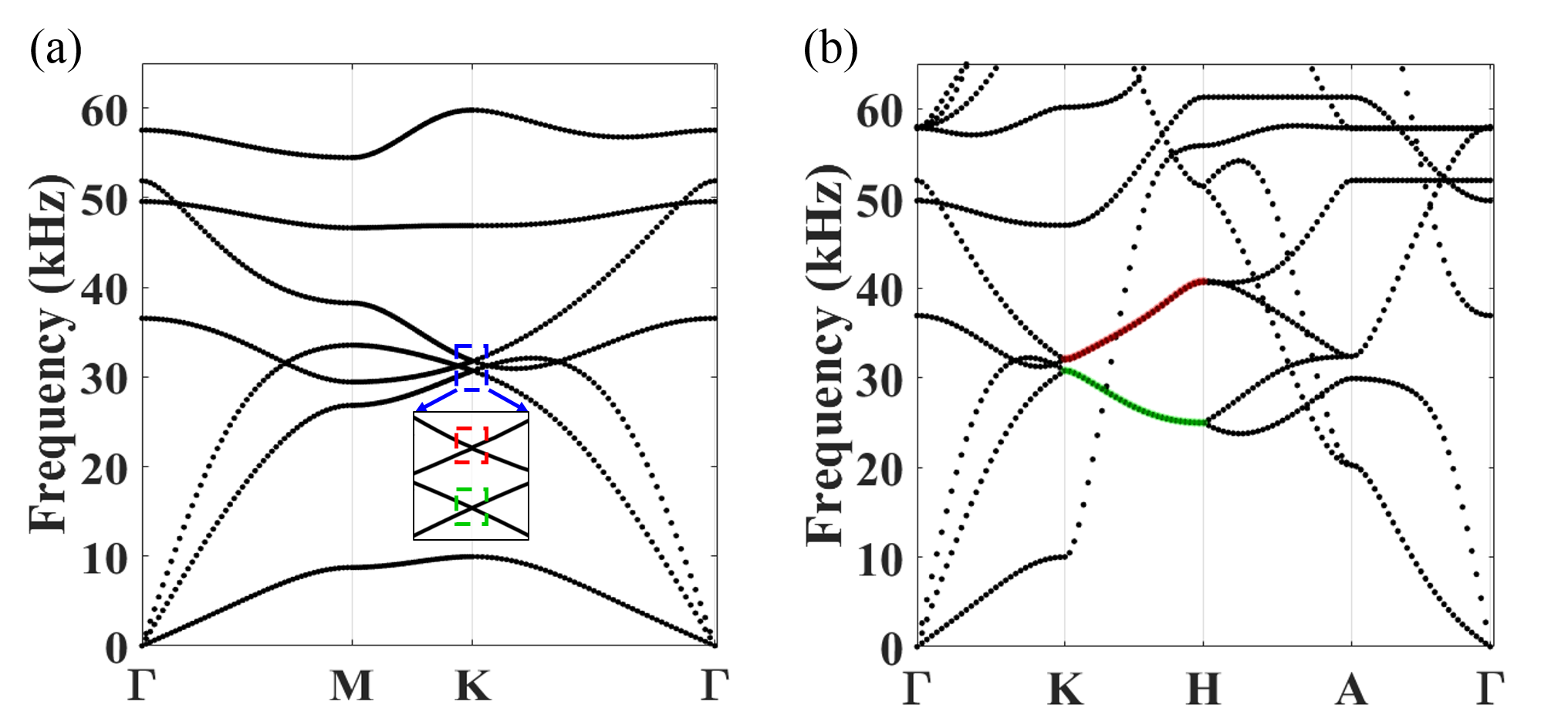}
	\caption{\label{fig:2} (a) Dispersion curves of the 3D lattice based on the unit cell shown in Fig.~\ref{fig:1}. (a) the dispersion along the 2D BZ in the $k_x\mhyphen k_y$ plane at $k_z=0$. The region marked in blue is magnified in the inset to show the presence of two Dirac points (marked by the red and green boxes). These degeneracies are protected by the symmetry of the triangular lattice in the $x\mhyphen y$ plane. (b) Dispersion curves plotted along the 2D BZ in the $k_x\mhyphen k_z$ plane at $k_y=0$. The nodal degeneracies at the corner of the BZ (marked in Fig.~\ref{fig:2} (a)) exist for all values of $k_z$, thereby resulting in line degeneracies along the K-H directions indicated by the red and green lines.}
\end{figure*}

The dispersion curves along the boundary of the 2D BZ in the $k_x\mhyphen k_y$ plane at $k_z=0$, and in the $k_x\mhyphen k_z$ plane at $k_y=0$ have been calculated using the commercial finite element (FE) package COMSOL Multiphysics and are shown in Fig.~\ref{fig:2}. As visible in Fig.~\ref{fig:2} (a), four modes intersect linearly at a frequency of approximately 40 kHz and give rise to two degenerate points at \textbf{K}. These degenerate points are indicated by the red and green boxes in the inset of Fig.~\ref{fig:2} (a). 

% Figure 2

The resulting degenerate nodes are Dirac points which are protected by the lattice symmetry of the hexagonal (graphene-like) arrangement of vertical cylinders. As previously mentioned, this degeneracy exists at all $k_z$, hence resulting in the line degeneracies marked in red and green colors in Fig.~\ref{fig:2} (b). This behavior is a direct consequence of both $P$ and $T$-symmetries being preserved. Breaking  $P$-symmetry would result in lifting the line degeneracy along the $k_z$ direction hence resulting in the formation of a Weyl points.

\subsection{\textit{P}-symmetry breaking and Weyl points}
% Figure 3
\begin{figure}[ht]
	\centering
	\includegraphics[scale=0.6]{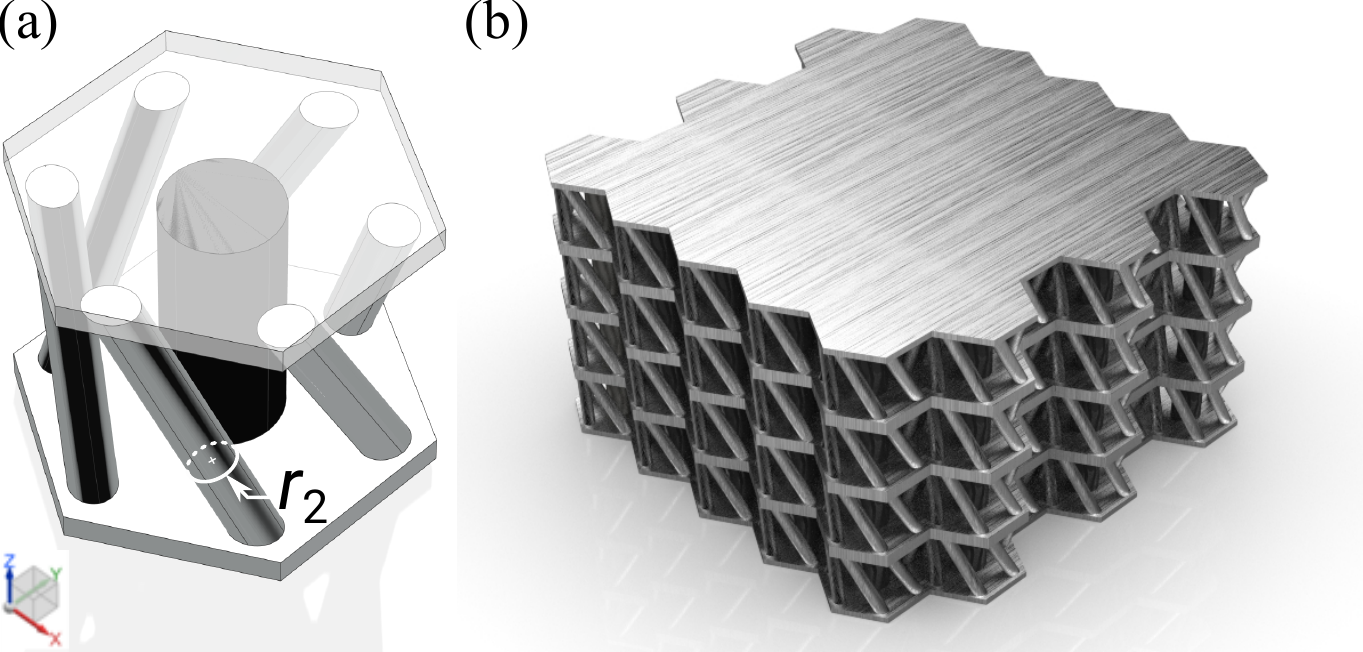}
	\caption{\label{fig:3} (a) Schematics of the unit cell with slanted cylinders. The cylinders have radius $r_2$ and are made of aluminum. The resulting unit cell does not preserve $P$-symmetry and mirror symmetry. Periodic boundary conditions are used on the boundaries to create a 3D lattice.
	(b) Rendered view of a bulk piece of the 3D lattice.}
\end{figure}
% Figure 4
\begin{figure*}
	\centering
	\includegraphics[trim={0cm 0cm 0cm 0cm},clip=true,scale=0.32]{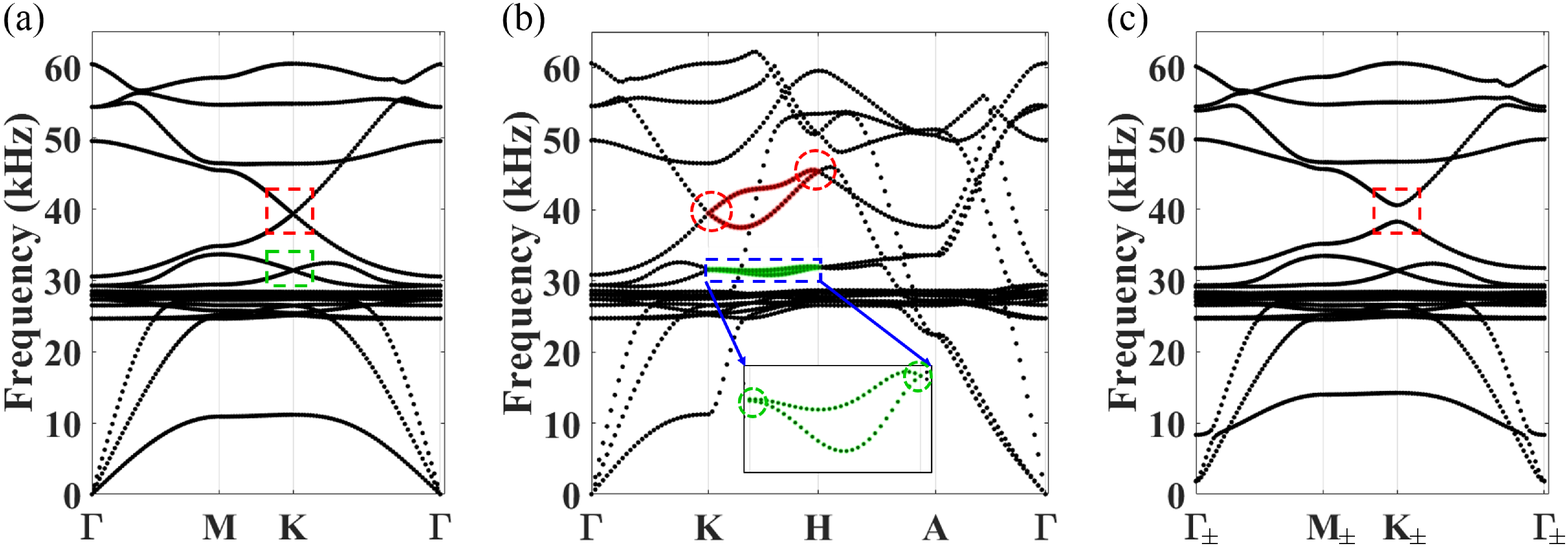}
	\caption{\label{fig:4} Dispersion curves for the 3D lattice with broken $P$-symmetry. (a) Dispersion along the 2D BZ in the $k_x\mhyphen k_y$ plane at $k_z=0$. The nodal degeneracies from Fig.~\ref{fig:2} (a) are still present, despite a slight shift in frequency. (b) Dispersion curves of the same 3D lattice along the 2D BZ in $k_x\mhyphen k_z$ plane at $k_y=0$. The line degeneracies in Fig.~\ref{fig:2} (b) are lifted (magnified image of the green modes is shown in the inset). The resulting nodal degeneracies at K and H are Weyl points, marked by red and green circles. (c) Dispersion curves along the 2D BZ in the $k_x\mhyphen k_y$ plane when $k_z=\frac{0.1\pi}{a_z}$. The degeneracy indicated by the red box in Fig.~\ref{fig:4} (a) is now lifted, due to the nonzero value of $k_z$, hence leaving behind a topological bandgap. The degeneracy marked by the green box in Fig.~\ref{fig:4} (a) is also lifted due to the nonzero value of $k_z$, which opens a second (although narrower) bandgap.}
\end{figure*}
In order to break $P$-symmetry, a chiral coupling can be introduced by means of straight slanted cylinders connecting top and bottom layers in proximity of the vertices of the hexagonal layer. The resulting unit cell is shown in Fig.~\ref{fig:3} while its 3D BZ remains the same as the one shown in Fig.~\ref{fig:1} (c). The slanted cylinders are made of aluminum (as the plates) and have a radius of $r_2=2$ mm. The dispersion curves along the boundary of 2D BZ in the $k_x$-$k_y$ plane at $k_z=0$ and in the $k_x$-$k_z$ plane at $k_y=0$ are calculated again and shown in Fig.~\ref{fig:4}. The degenerate points resulting from the linear intersection of two modes in Fig.~\ref{fig:2} (a) are also present in Fig.~\ref{fig:4} (a) (marked by the same red and green boxes). However, the line degeneracies previously observed in Fig.~\ref{fig:2} (b) are now lifted, due to $P$-breaking, and reduced to nodal degeneracies at the high-symmetry points \textbf{K} and \textbf{H} (see red and green circles Fig.~\ref{fig:4} (b)). These degenaracies are the Weyl points that are formed by the linear intersection of the two (initially degenerate) modes along all three directions. Fig.~\ref{fig:4} (a) shows the linear intersection in the $k_x$-$k_y$ plane while Fig.~\ref{fig:4} (b) shows the linear intersection in $k_x$-$k_z$ plane.

As a result, the degeneracy marked in Fig.~\ref{fig:4} (a) is lifted for nonzero values of $k_z$ (with $k_z \neq n \pi / a_z$ for integer $n$) hence giving rise to a topological bandgap as shown in Fig.~\ref{fig:4} (c). The band structure shown in Fig.~\ref{fig:4} (c) corresponds to $k_z=\pm\frac{0.1\pi}{a_z}$, where $a_z=h_1+2h_0=24$ mm is the total height of unit cell. The subscripts $\pm$ of the high symmetry points' labels indicate $k_z=\pm\frac{0.1\pi}{a_z}$, as also depicted in Fig.~\ref{fig:1} (c).
The bandgap size varies with $k_z$ as shown in Fig.~\ref{fig:4} (b). Thus, $k_z$ can be treated as a parameter that controls either the opening or closing of the topologically nontrivial bandgap, provided that the system remains periodic along the $z$-direction.   

\section{\textbf{Ab initio} calculations of the topological properties}
\label{sec3}
\subsection{Topological charge}
In order to assess the topological significance of the degeneracies identified in the chiral 3D lattice and, consequently, the existence of Weyl points, this section presents a numerical investigation into the calculation of the topological charge. The charge can be calculated by integrating the Berry curvature flux on a 2D manifold $S$ enclosing the Weyl point in the reciprocal space, that is
\begin{equation}
\label{eq:C}
C=\frac{1}{2\pi}\oint_S \bm{\Omega}(\mathbf{k})\cdot d\mathbf{S}
\end{equation}
where $C$ is the topological monopole charge, i.e., the Chern number, $d\mathbf S$ is the vector surface element aligned with its local normal direction, and $\bm{\Omega}$ is the Berry curvature (vector field) given by
\begin{equation}
\label{eq:BC}
\bm{\Omega}(\mathbf{k})=\nabla _{\mathbf{k}} \times\langle \mathbf{u}(\mathbf{k})|i\nabla _{\mathbf{k}}|\mathbf{u}(\mathbf{k})\rangle
\end{equation}
where $\mathbf{k}$ is the wavevector, and $\mathbf{u}_n(\mathbf{k})$ is the displacement eigenstate as a function of $\mathbf{k}$.
\begin{figure}[ht]
	\centering
	\includegraphics[scale=0.9]{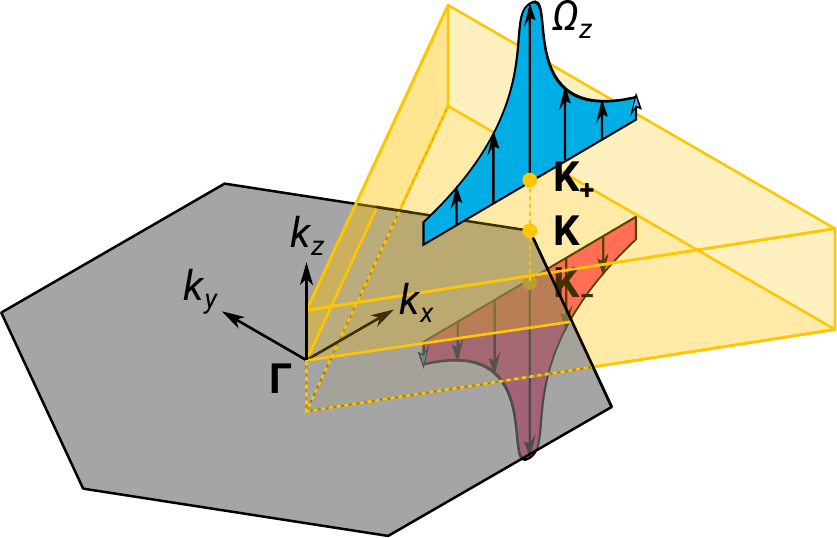}
	\caption{\label{fig:charge} The topological charge at the K point can be calculated by integrating the outward Berry curvature flux of a 2D manifold enclosing the K point. In this case, the surfaces of a triangular prism are chosen.}
\end{figure}
Consider the Weyl point at \textbf{K}, at a frequency of approximately 40 kHz, as indicated by the red dashed box in Fig.~\ref{fig:4} (a). This point corresponds to the intersection of the 16$^{th}$ and the 17$^{th}$ bands. To calculate the corresponding topological charge, we can integrate the Berry curvature flux threading the outer surface area (on the top, bottom, and side faces) of a fictitious prismatic manifold enclosing the \textbf{K} point (see Fig.~\ref{fig:charge}). We anticipate, and show later, that the flux on the side faces is zero if a certain triangular prism is chosen as prototypical 2D manifold.

\begin{figure}[ht]
	\centering
	\includegraphics[scale=0.54]{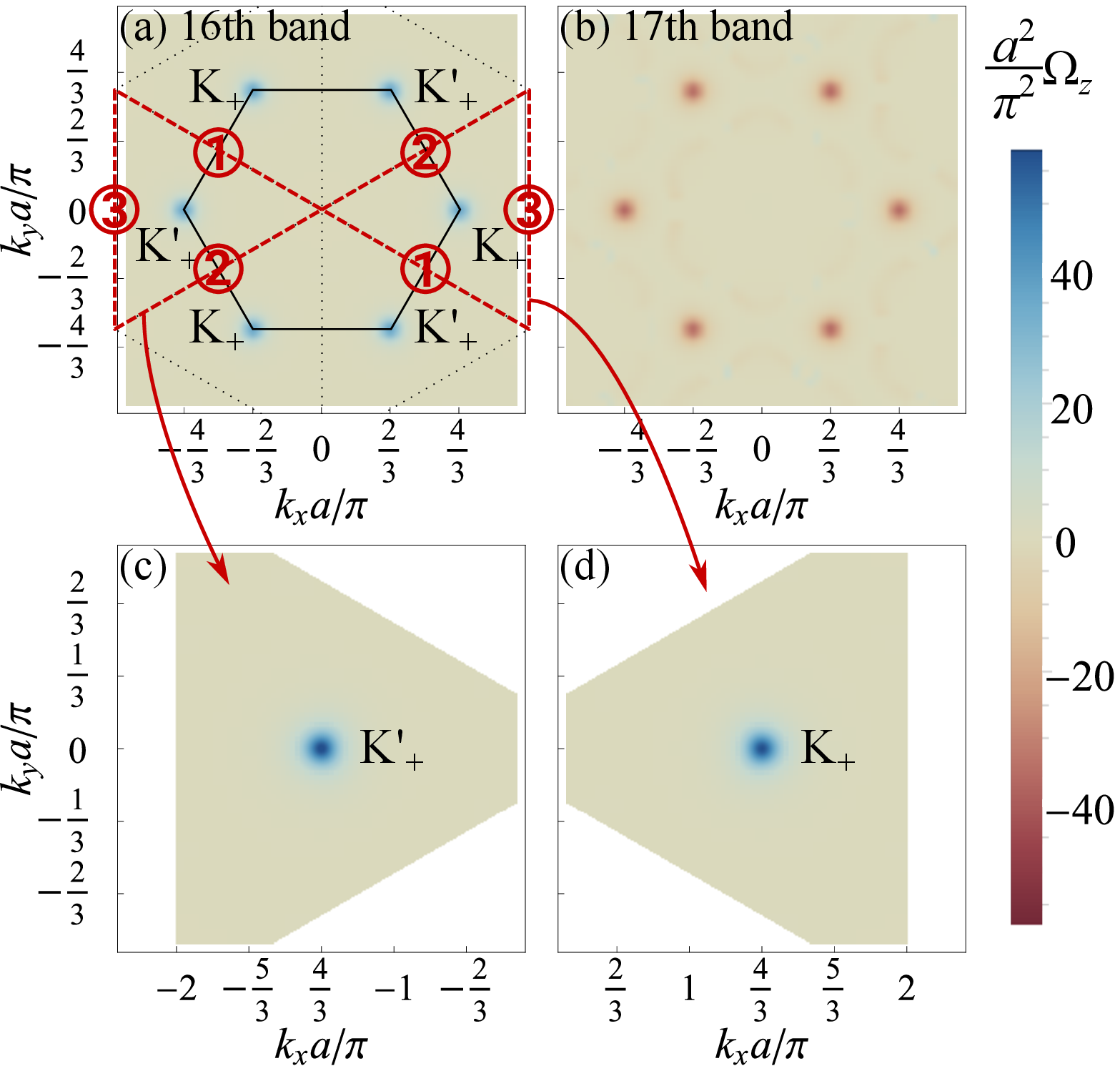}
	\caption{\label{fig:BC} (a,b) The $z$-Berry curvature distribution of the $16^{th}$ and $17^{th}$ bands, respectively, on the $k_z=0.1\pi /a_z$ plane. The two red dashed triangles in (a) indicate the top faces of the chosen prismatic manifolds enclosing the Weyl points, K or $\textrm{K}'$ points, respectively. The total integral of the Berry curvature flux on the vertical side walls \ding{172}, \ding{173} and \ding{174} of the prism vanishes due to the $C_6$ and translation symmetry. (c,d) Zoomed-in view of the Berry curvature corresponding to the two triangular areas in (a). Each yields the outward flux integral of $\pi$, which shows the Weyl points K and $\textrm{K}'$ are of $+1$ Chern number.
	}
\end{figure}

Without loss of generality, we can select $k_z=\pm\frac{0.1\pi}{a_z}$ as the top and bottom faces of the prism. At these planes, we have \textbf{K$_+$} and \textbf{K$_-$} with the same $(k_x, k_y)$ coordinate as the \textbf{K} point.
The $z$-component of the Berry curvature on the top plane is then calculated using Eq.~\ref{eq:BC} (see Fig.~\ref{fig:BC} (a) and (b)) for the $16^{th}$ and $17^{th}$ bands, respectively.
Note that, at the two distinct points (i.e. they cannot be transformed into one another by translation symmetry operations), \textbf{K$_+$} and $\textrm{\textbf{K}}'_+$ points exhibit the same Berry curvature pattern. This is a direct result of the $C_6$ symmetry of the lattice. Therefore, if the 2D manifold is chosen to be the triangular prism with the right red dashed triangle in Fig.~\ref{fig:BC} (a) as its top face, we can assure that the flux integral on the side faces of the prism is zero due to symmetry in the reciprocal space. In fact, the $C_6$ symmetry guarantees that the outward flux on the side face \ding{172} equals the inward flux on the side face \ding{173}; the same cyclic symmetry along with the translation symmetry (from the left to the right, for example) forces the flux integral on the entire side surface \ding{174} to vanish.

\begin{figure}[ht]
	\centering
	\includegraphics[scale=0.33]{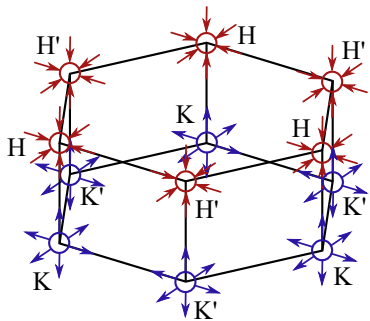}
	\caption{\label{fig:charges}  Schematic illustration of the Weyl points within the $k_z > 0$ half BZ and their corresponding topological monopole nature (either as a sink or a source).}
\end{figure}

It follows that we only need to integrate the Berry curvature flux on the top and bottom faces. The bottom face which is around the \textbf{K$_-$} point, is actually the time reversed counter-part of the triangular face around the $\textrm{\textbf{K}}'_+$ point (inversion of $(k_x, k_y, k_z)$) as shown in the left red dashed triangle in Fig.~\ref{fig:BC} (a). $T$-symmetry requires that $\bm{\Omega}(-\mathbf{k})=-\bm{\Omega}(\mathbf{k})$ which results in the $z$-component of the Berry curvature on the bottom face to have the same strength but opposite sign. Also, given that the $d\mathbf{S}$ element on the bottom face is oriented in the $-k_z$ direction, it still contributes a positive outward flux. As a result, the topological charge is the sum of the integrals of the Berry curvature calculated over the two red dashed triangles in Fig.~\ref{fig:BC} (a). While solving for the Bloch eigenmodes, we performed a parametric analysis across the $k_x$ and $k_y$ grid points by using an adaptive resolution (finer near $\textrm{\textbf{K}}'_+$ point). The Berry curvature was calculated based on Eq.~\ref{eq:BC} using a finite difference formulation on the two triangular regions, and the results are shown in Fig.~\ref{fig:BC} (c) and (d). The numerical integrals yield 3.1561 and 3.1479, respectively, hence confirming the monopole charge of $+1$ (with only $+0.3\%$ error).
Using a very similar numerical approach, the topological charge at the \textbf{H} point can be found to be $-1$. This result is in agreement with the $T$-symmetric nature of the Weyl points which requires the appearance of the degeneracies in pairs of points with opposite charge. Fig.~\ref{fig:charges} shows all the Weyl points within the half BZ (in the $k_z$ direction) and their corresponding topological monopole nature (either sink or source).
On the $k_z=0$ plane, there are two Weyl points (\textbf{K} and $\mathrm{\textbf{K}}'$) both with charge $+1$ (that is sources of Berry curvature), while on the $k_z=\pi/a_z$ plane, each of the two Weyl points (\textbf{H} and $\mathrm{\textbf{H}}'$) are of charge $-1$ (that is sinks of Berry curvature).

\subsection{Bandgap-Chern number}
An alternative way to obtain the topological charge of the Weyl point is to use a topological invariant called the bandgap-Chern number $C_{g}$.
This invariant is obtained by summing the Chern numbers of all the bands below the gap of interest. In the following, we consider $k_z$ as a parameter and focus on the bandgap that forms in the $k_x\mhyphen k_y$ plane. $C_g$ remains constant with varying $k_z$ except at \textbf{K} and \textbf{H} where the bandgap closes and reopens and $C_g$ experiences a sudden change in its value. This change is equal to the topological charge of the Weyl points on that plane because Weyl points act as either sources or sinks of Berry curvature flux \cite{takahashi2018circularly}. This principle can be used to calculate the charge of the Weyl points in the band structure. Particularly, we focus on the degeneracies occurring at approximately 40 kHz. %\FS{You are making the point of using the data from all the bands. If you make this assumption, doesn't it defeat the purpose? I mean, you're not showing it is true.} \ed{[Then I put it in a different way that I am not assuming but I found it to be. (do you think if I should put the plots of Berry curvature from 1st to 15th modes, like in the appendix? It should be visually evident even without numerical integration)]}
It is found that the Berry curvature integrals of the first 15 bands cancel each others, therefore $C_g$ reduces to the integral of Berry curvature in the 2D BZ for the 16$^{th}$ mode only.

\begin{figure}[ht]
	\centering
	\includegraphics[scale=0.6]{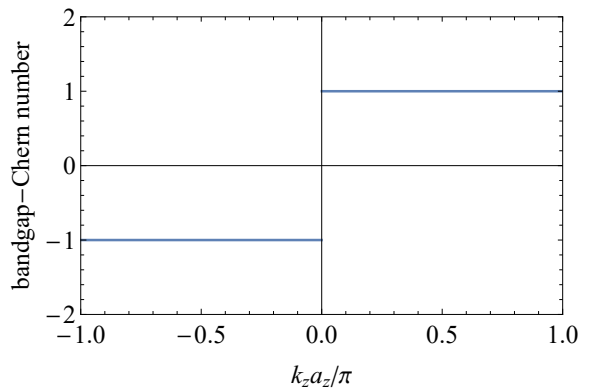}
	\caption{\label{fig:BGC} The bandgap-Chern number $C_g$ associated with the gap around 40 kHz as a function of $k_z$. As the bandgap closes and reopens at $k_z=0$, $C_g$ experiences a $+2$ discontinuous jump in its value, which indicates the sum of topological monopole charges on the $k_z=0$ plane.}
\end{figure}

Figure~\ref{fig:BGC} shows the plot of $C_g$ as a function of $k_z$ for the Weyl points at 40kHz. It clearly shows that $C_g$ has a value of $-1$ for $k_z<0$ and $+1$ for $k_z>0$ which results in a net difference $\Delta C_g=+2$. This is consistent with the previous analysis that showed the $k_z=0$ plane to host two Weyl points of equal charge at \textbf{K} and $\mathrm{\textbf{K}}'$ due to intact $T$-symmetry. The two Weyl points at \textbf{K} and $\mathrm{\textbf{K}}'$ therefore have a topological charge of $+1$ each. In the same figure, there is also a change in the value of $C_g$ at $k_z=\pm\frac{\pi}{a_z}$. This indicates that the charge of the Weyl point at \textbf{H} ($\mathrm{\textbf{H}}'$) is $-1$. Once again, this is expected and consistent with the fact that Weyl points always occur in pairs with opposite topological charge. The nonzero value of charge also confirms that the nodal degenerate points under consideration are in fact Weyl points with nontrivial topological significance. In light of the bulk-surface correspondence \cite{hasan},
a nonzero $C_g$ indicates that topological surface modes can exist within the target bandgap while the magnitude of $C_g$ indicates how many surface modes should be expected. In our specific example,
$C_g =\pm 1$ for $k_z \in (0, \pi/a)$ or $k_z \in (-\pi/a,0)$, respectively. Compared with trivial materials (such as vacuum), there is a $k_z$-locked net difference of $\pm 1$ in $C_g$. Therefore, one surface mode should be expected either in case of positive or negative $k_z$. The corresponding surface states propagate unidirectionally either in the clockwise or counter-clockwise direction along the structure border (depending on the sign of $C_g$), as shown later on in the numerical simulations. This result further confirms that the selected geometry leads to the formation of Weyl points and it is capable of supporting topological surface states.
%

%\FS{Based on our previous comments below, see if you agree with the new text} 
We note that the nonzero Chern number and the Berry curvature distribution for, as an example, $k_z=0.1\pi/a_z$ (Fig.~\ref{fig:BC}) is reminiscent of 2D topological materials manifesting quantum Hall effect (QHE) \cite{raghu2008analogs}. In QHE materials,
%\FS{Are you trying to make a parallel between the propagation along z to be seen as broken T symmetry?} \ed{[exactly]} 
$T$-symmetry is broken due to the use of an external field. In the present material, the chiral structure breaks the $z$-mirror symmetry therefore a mode propagating in the $z-$direction in the infinite lattice is expected to exhibit $k_z$-locked unidirectional surface states. In fact, at steady state, the role of temporal and the spatial (in this case $z$) variables can be interchanged.
A similar idea was also presented in studies of 3D Floquet topological insulators \cite{rechtsman2013photonic}.
This latter aspect will be further clarified in the next section.

\section{Supercell dispersion and surface modes}
\label{sec4}
\begin{figure*}[ht]
	\centering
	\includegraphics[trim={0cm 0cm 0cm 0cm},clip=true,scale=0.56]{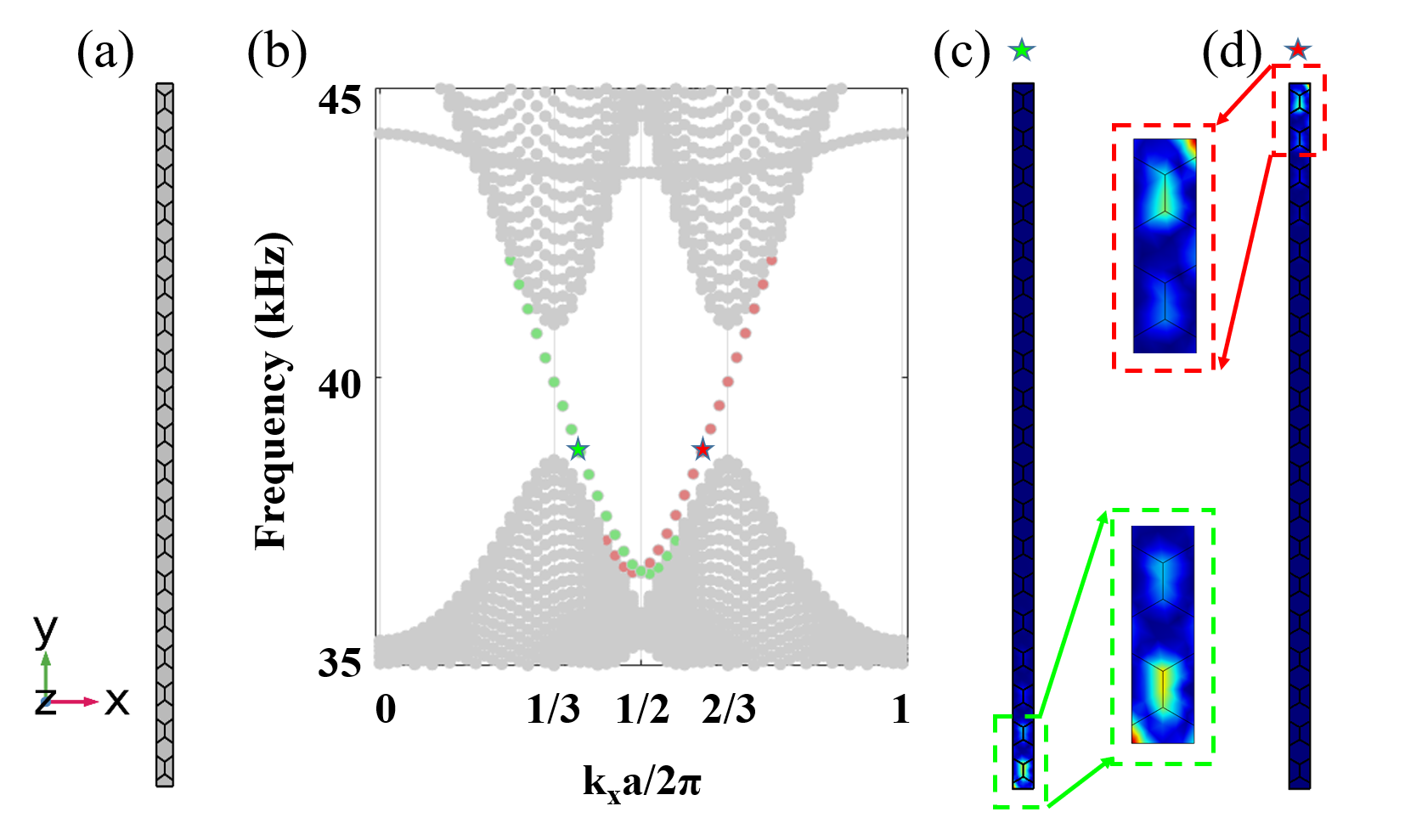}
	\caption{\label{fig:7} (a) Supercell domain having finite size in the $y$-direction (38 units) and infinite size (via the application of periodic boundary conditions) in the $x$- and $z$-directions. (b) Dispersion curves corresponding to the supercell at $k_z=\frac{0.1\pi}{a_z}$ and in the neighborhood of the bandgap marked by red box in Fig.~\ref{fig:4} (c). Topological surface modes exist in the bandgap and are indicated in red and green color. The bulk modes are indicated in grey color. The green colored modes occur at the lower edge of the supercell while the red colored modes occur at the upper edge as indicated in Fig.~\ref{fig:7} (c) and Fig.~\ref{fig:7} (d), respectively. The eigenvectors corresponding to the points indicated by (c) the green star, and (d) the red star in the dispersion curves and plotted on the supercell. }
\end{figure*}

Weyl points with opposite topological charges are connected by nontrivial topological surface states that exist only on the boundary of the medium. In this section, we consider the physical response of a periodic supercell made out of the 3D unit cell described above. Consider a supercell having a finite dimension of 38 units in the $y$ direction and infinite dimensions (obtained by means of periodic boundary conditions) in the $x$ and $z$ directions. Fig.~\ref{fig:7} (a) shows the actual domain used for the numerical calculations.
% Figure 7

We focus the following analysis on the topological bandgap that develops at the approximate frequency of 40kHz and that is marked by the red box in Fig.~\ref{fig:4} (c). Fig.~\ref{fig:7} (b) shows the dispersion curves for the supercell at $k_z=\frac{0.1\pi}{a_z}$. It is evident from the visual inspection of the dispersion that the supercell admits two additional modes, the surface modes, that exist within the topological bandgap. Bulk modes are in grey color while the surface modes are indicated in green and red. The green modes correspond to the lower edge of the supercell while red modes correspond to upper edge. This correspondence is also reflected in the plot of the eigenvectors on the supercell shown in Fig.~\ref{fig:7} (c) and \ref{fig:7} (d). The dispersion of the surface modes corresponding to the upper edge has a positive slope (i.e. positive group velocity), which means that these modes are expected to travel in the positive $x$-direction. Similarly, the surface modes defined on the lower edge have negative slope and therefore are expected to travel in the negative $x$-direction. 

\begin{figure*}[ht]
	\centering
	\includegraphics[trim={0cm 0cm 0cm 0cm},clip=true,scale=0.55]{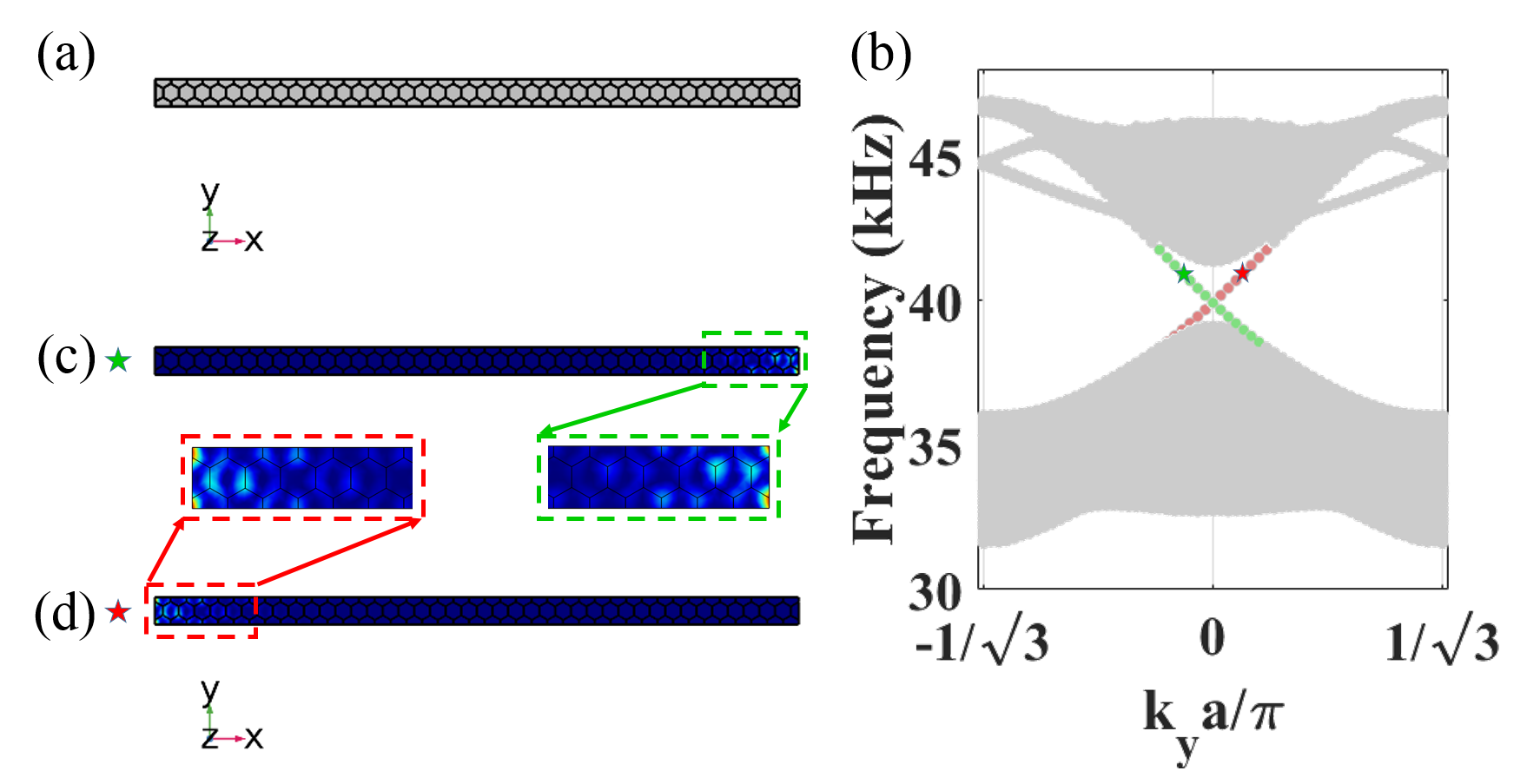}
	\caption{\label{fig:8} (a) Supercell domain having finite size in the $x-$direction (40 units) and infinite size (via the application of periodic boundary conditions) in the $y$- and $z$-directions. (b) Dispersion curves corresponding to the supercell at $k_z=\frac{0.1\pi}{a_z}$ around the bandgap marked by red box in Fig.~\ref{fig:4} (c). Topological surface modes exist in the bandgap and are indicated in red and green color, while the bulk modes are indicated in grey color. The green colored modes occur at the right edge of supercell while red colored modes occur at the left edge of the supercell as indicated in Fig.~\ref{fig:8} (c) and Fig.~\ref{fig:8} (d) respectively. The eigenvectors corresponding to the points indicated by (c) the green star (d) the red star in the dispersion curves and plotted on the supercell. }
\end{figure*}

Further, it can be observed that the supercell in Fig.~\ref{fig:7} is obtained by terminating the triangular lattice along the zigzag edge. However, triangular lattices exhibit also another edge type named the armchair edge. We explored the possible propagation and existence of topological modes also on this edge type. For this purpose, a finite size supercell terminated along the $x$-direction was developed, as shown in Fig.~\ref{fig:8} (a). The supercell extended indefinitely in the $y$ and $z$ directions (periodic boundary conditions were used to simulate the infinite size). The corresponding band structure at $k_z=\frac{0.1\pi}{a_z}$ is shown in Fig.~\ref{fig:8} (b). The bulk modes are marked in grey color, the surface modes corresponding to the right edge of the supercell are indicated in green color, and those corresponding to the left edge are indicated in red color. The eigenvectors corresponding to the surface states on both the right and left edges are shown in Fig.~\ref{fig:8} (c) and \ref{fig:8} (d), respectively. Similar to the zigzag edge, the dispersion of the surface modes corresponding to the left edge has a positive slope (i.e. positive group velocity), which means that these modes are expected to travel in the positive $y$-direction. The surface modes defined on the right edge have negative slope and therefore are expected to travel in the negative $y$-direction. Therefore, if we assemble an extended 2D domain in the plane $x$-$y$ we would expect to see the surface mode to propagate in a clockwise direction along the boundary of the lattice. 

For $k_z=-\frac{0.1\pi}{a_z}$, the dispersion curves remain unaltered but the direction of propagation of the surface modes is reversed. The surface mode corresponding to the upper edge acquires a negative group velocity (propagation in the negative $x$-direction) while the surface mode corresponding to the lower edge acquires a positive group velocity (propagation in the positive $x$-direction). Similarly, the surface mode corresponding to the left edge acquires a negative group velocity (propagation in the negative $y$-direction) while the surface mode corresponding to the right edge acquires a positive group velocity (propagation in the positive $y$-direction). An extended 2D domain in the $x$-$y$ plane would then show a wave propagating in the anti-clockwise sense along the boundaries. 

The combined behavior illustrated above is well consistent with those of dynamical systems characterized by Weyl points and corresponding topologically nontrivial modes. Results also suggest that, upon injecting a wave into the system, the direction of propagation can be controlled by properly choosing the sign of $k_z$. This latter aspect will be further clarified by means of full field simulations in the next section.

% Figure 8

\begin{figure*}[ht]
	\centering
	\includegraphics[trim={0cm 0cm 0cm 0cm},clip=true,scale=0.36]{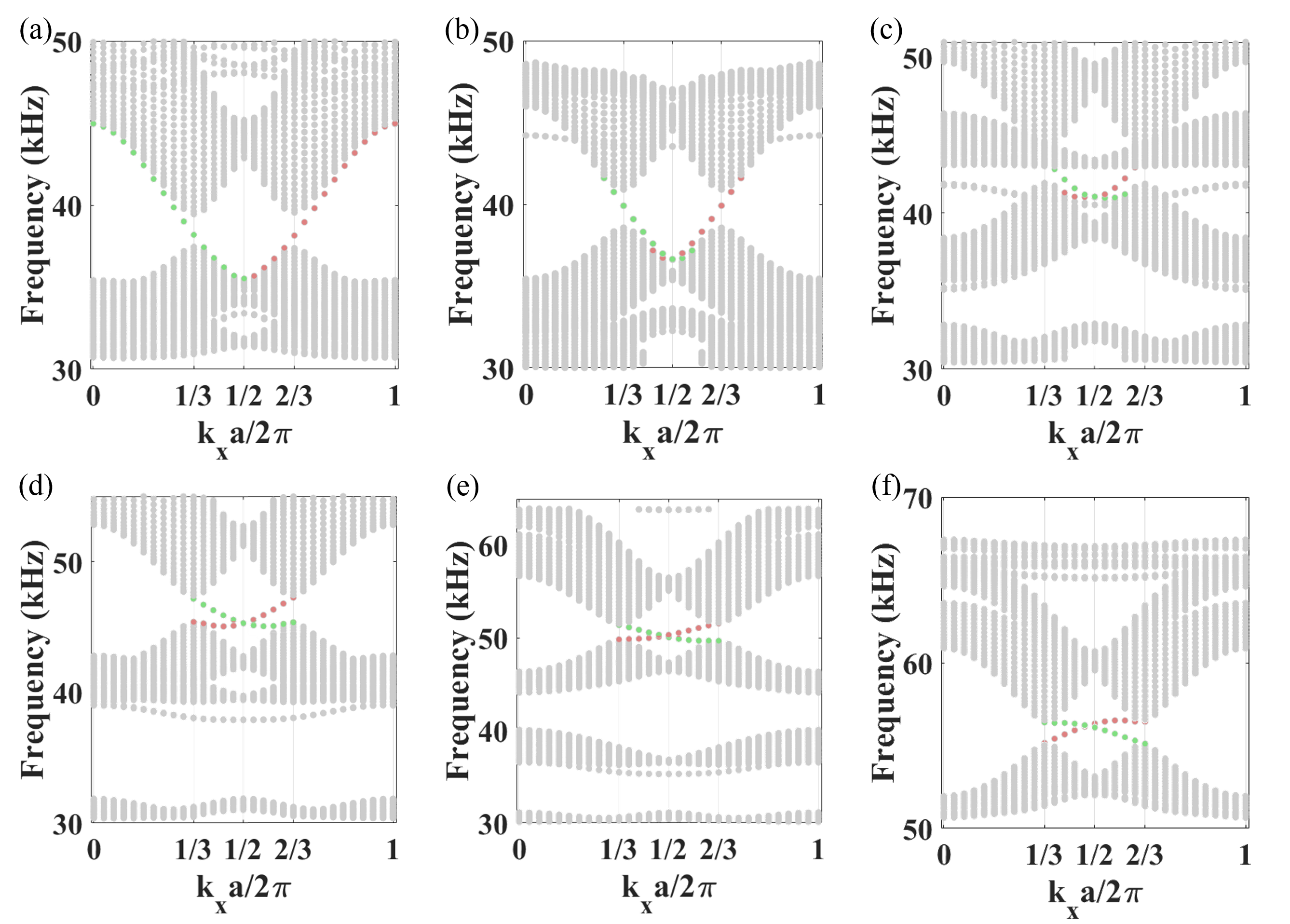}
	\caption{\label{fig:9} Dispersion curves of the supercell shown in Fig.~\ref{fig:7}(a) for different values of the radius of the center cylinder. Values of $r_1$ equal to (a) 4mm (b) 5mm (c) 6mm (d) 7mm (e) 8mm (e) 9mm are considered. It is seen that the group velocity of the topological surface modes decreases with increasing $r_1$ and that the surface states evolve from a single-valley to an inter-valley dominated behavior. }
\end{figure*}

Another interesting feature of the proposed geometry is that the group velocity of the surface modes can be controlled simply by changing the radius of the center vertical cylinder $r_1$. Fig.~\ref{fig:9} shows the dispersion curve of the supercell (a) for various values of $r_1$. The group velocity of the surface modes decreases with increasing radius $r_1$ of the center cylinder.
This reduction of the group velocity can be explained by observing that by increasing $r_1$ the homogenized mass density of the lattice increases, but without significantly increasing the rigidity. Hence, given that the eigenmode does not involve significant compressive or bending motion of the center cylinder, the surface modes tend to slow down. By tuning the radius $r_1$, the surface states can gradually evolve from the single valley mechanism (similar to the 2D  model discussed by Raghu\cite{raghu2008analogs}) to the intervalley mechanisms (similar to the model discussed in Kane\cite{Z2}). In either case, the unidirectional gapless surface states connecting the lower and upper bulk bands are protected by the topological nature of the band structure and are guaranteed to be present.
Note that, while a similar variation in the group velocity of the topological edge modes was earlier observed in 2D chiral materials \cite{rechtsman2013photonic}, this tuning ability of the surface states in 3D lattice was never observed and documented in earlier studies.
% Figure 9

\section{Full field numerical simulations}
\label{sec5}

\begin{figure*}[ht]
	\centering
	\includegraphics[trim={0cm 0cm 0cm 0cm},clip=true,scale=0.43]{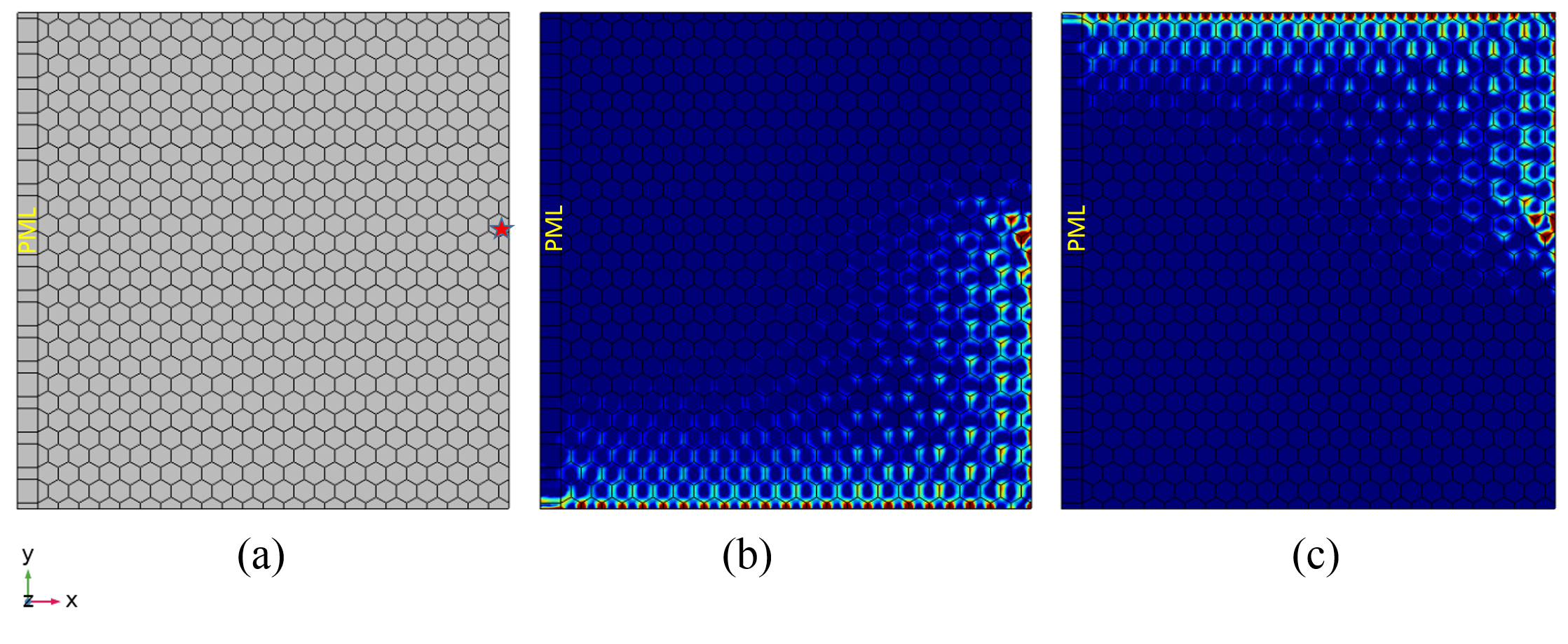}
	\caption{\label{fig:10} (a) Domain used for full field simulations. The domain is finite in size along the $x$- and $y$-directions but it employs periodic boundary conditions along the $z$-direction. A PML boundary was applied along the left edge of the domain to absorb the incoming wave and to facilitate the visualization of the direction of propagation of the surface mode during steady state analyses. The wave was injected by means of a harmonic point force applied at the location marked by the red star at a frequency of 40 kHz. (b) Eigenstate showing the propagation of the surface state when excited with $k_z=\frac{0.1\pi}{a_z}$. (c) Eigenstate showing the propagation of the surface state when excited with $k_z=\frac{-0.1\pi}{a_z}$. These results clearly show the unidirectional nature of the topological surface modes.}
\end{figure*}

Based on the 3D unit cell developed above, a complete domain in the $x$-$y$ plane was also simulated in order to observe both the direction of propagation and the scattering behavior in presence of defects. The domain used is shown in Fig.~\ref{fig:10} (a). The domain is rectangular and finite along both the $x$- and $y$-directions. Both edges were set to free boundaries. The boundaries normal to the $z$-axis were assigned periodic boundary conditions so to render the lattice infinite in this direction. Perfectly matched layers (PML) were used at the left edge to absorb the incoming wave and allow a clear assessment of the direction of propagation of the surface state when performing a steady state analysis (i.e. it avoids the surface mode to propagate all around the boundary and get back to the source).

% Figure 10

The domain was excited by a harmonic point force acting in the $z$-direction at the location indicated by the red star marker in Fig.~\ref{fig:10} (a). A harmonic excitation at a frequency of 40 kHz was selected because right within the target topological bandgap where surface states exist. At $k_z=\frac{0.1\pi}{a_z}$, the source excites the surface state with positive group velocity which is observed propagating in the clockwise direction (see Fig.~\ref{fig:10} (b)). At $k_z=-\frac{0.1\pi}{a_z}$, the situation is inverted and the source excites an counter-clockwise propagating surface state (Fig.~\ref{fig:10} (c)). These results confirm previous observations concerning the one-way propagation property of the surface states. 

These results also allow an additional observation on the back scattering immune properties of the surface states. As the wave propagates beyond the corner of the rectangular domain, the wave amplitude on both sides is exactly comparable. This suggests that there are no significant reflections taking place at the sharp corner and the wave is capable of propagating unidirectionally.

It should be noted that while the states propagate with the same intensity along the $x$- and $y$-aligned edges, they are associated with different eigenstates. This can be clearly observed in Fig.~\ref{fig:10} (b) and \ref{fig:10} (c), and it is due to the different structure of the edges (i.e. zigzag versus armchair). %This difference in the kind of edges that can be formed from graphene-like unit cell, differentiates the different mode shapes that can occur at the boundary.

% Figure 11
\begin{figure*}[ht]
	\centering
	\includegraphics[trim={0cm 0cm 0cm 0cm},clip=true,scale=0.5]{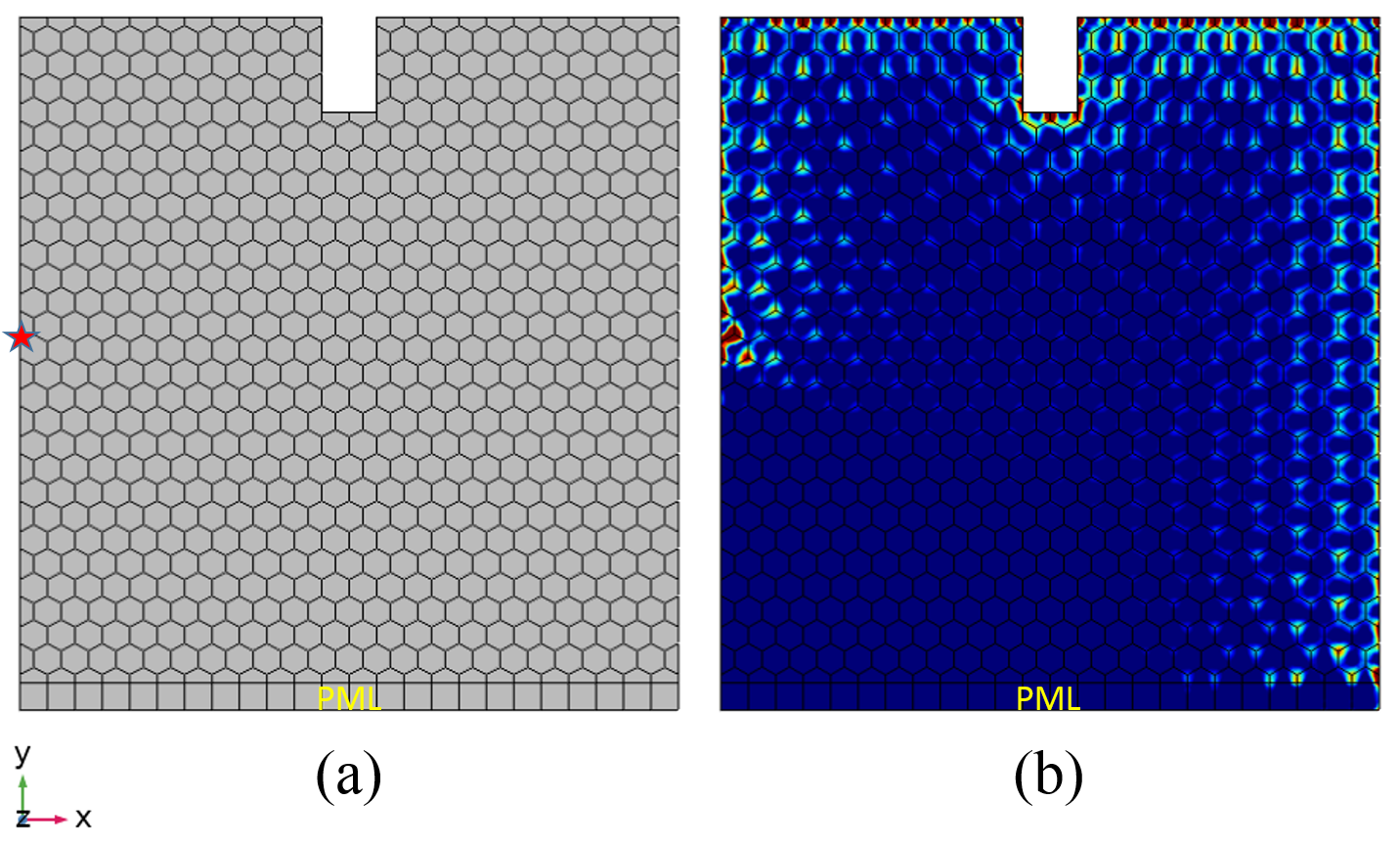}
	\caption{\label{fig:11} (a) Full field simulations on a domain with an edge defect represented by a cut-out at the top boundary. The domain is finite along the $x$- and $y$-directions and it has periodic boundary conditions along the $z$-direction. PMLs were used along the bottom boundary to absorb the incoming wave and facilitate the observation of unidirectionally propagating states. Harmonic point load applied at the location indicated by red star at a frequency of 40 kHz. (b) Response showing the propagation of the surface wave when excited with $k_z=\frac{0.1\pi}{a_z}$. The wave clearly travels unidirectionally from the source and can efficiently propagate around the defect with no appreciable back-scattering.}
\end{figure*}

The immunity to back scattering was also assessed by introducing a defect in the rectangular domain represented by a cutout on the top edge, as shown in Fig.~\ref{fig:11} (a). The source was located on the left edge as indicated by the star marker. The excitation was kept identical to the previous simulation scenario.
By exciting at $k_z=\frac{0.1\pi}{a_z}$, a clockwise propagating surface state is clearly obtained as shown in Fig.~\ref{fig:11} (b). As the state encounters the defect on the top edge, it propagates around its perimeter and it continues along the original direction. Observing that the intensity of the wave amplitudes both before and after the defect are exactly comparable, and that there are no waves traveling back towards the source (reaching the section of the edge before the source), one can conclude that no appreciable back-scattering takes place due to the defect. The latter is a clear hallmark of topologically protected states.

\section{Conclusions}

Weyl points and topological surface modes in elastic solid media are still in their early stages of formulation and design. The very few implementations proposed to-date are based on discrete designs that are not conducive to use in practical applications, particularly in those requiring load-bearing materials. 
This study presented a fully continuous, load-bearing, elastic system capable of unidirectionally-propagating and topologically-protected surface states. The design of the 3D unit cell consisted in a layered prismatic lattice with hexagonal cross section in which the layers were spaced by solid cylindrical elements and by slanted circular beams connecting consecutive faces of the prismatic unit cell. The cylinders were used to set the necessary in-plane symmetry conditions and to provide high load carrying capacity. The slanted beams were used to achieve chirality and allowed achieving the necessary $P$-symmetry breaking conditions. Note that the layered structure resulted in a very feasible and practical design that has the potential to greatly facilitate the fabrication phase. 
The analysis of the lattice dynamics highlighted the existence of Weyl points following the breaking of the $z$-mirror-symmetry and the $P$-symmetry of the lattice. To gain insight into the mechanism leading to the formation of these degeneracy points, we evaluated the topological invariants using \textit{ab initio} calculations and without employing any further simplifications. Results were very well aligned with the expected quantized values.
The relationship between the surface states and the topological invariants was also elaborated upon from different perspectives. $k_z$-locked unidirectional propagating surface elastic states were predicted on the external surface of the 3D lattice and their existence was further confirmed by full field numerical simulations. Numerical results also confirmed the extreme robustness of these states against strong lattice defects.
This study may serve as a basis to develop structures having surface-elastic-wave-guiding capabilities for applications in fields such as vibration control, energy harvesting, structural health monitoring, and on-chip telecommunication signal processing.

\begin{acknowledgments}
The authors gratefully acknowledge the financial support of the National Science Foundation under Grant No. 1761423.
\end{acknowledgments}

\bibliography{Finaldraft}% Produces the bibliography via BibTeX.

\end{document}